\documentstyle[12pt]{article} 
\input psfig.sty

\def\Title#1{\begin{center} {\Large #1 } \end{center}}
\def\Author#1{\begin{center}{ \sc #1} \end{center}}
\def\Address#1{\begin{center}{ \it #1} \end{center}}

\def\doeack{\footnote{Work supported by the Department of Energy,
                     contract DE--AC03--76SF00515.}}
\def\SLAC{Stanford Linear Accelerator Center\\
    Stanford University, Stanford, California 94309 USA}

\newenvironment{Abstract}{\begin{quotation} \begin{center}
                       ABSTRACT
     \end{center}\bigskip  }{\end{quotation}}
\def\beq{\begin{equation}}
\def\eeq#1{\label{#1}\end{equation}}
\def\eeqn{\end{equation}}
\def\beqa{\begin{eqnarray}}
\def\eeqa#1{\label{#1}\end{eqnarray}}
\def\eeqan{\end{eqnarray}}

\def\Acknowledgements{\bigskip  \bigskip \begin{center} \begin{large}
             \bf ACKNOWLEDGEMENTS \end{large}\end{center}}

\def\Re{{\cal R \mskip-4mu \lower.1ex \hbox{\it e}\,}}
\def\Im{{\cal I \mskip-5mu \lower.1ex \hbox{\it m}\,}}
\def\nn{\noindent}
\def\ie{{\it i.e.}}
\def\eg{{\it e.g.}}

\def\etal{{\it et al.}}

\def\sub#1{_{\lower.25ex\hbox{$\scriptstyle#1$}}}
\def\sul#1{_{\kern-.1em#1}}
\def\sll#1{_{\kern-.2em#1}}  
\def\sbl#1{_{\kern-.1em\lower.25ex\hbox{$\scriptstyle#1$}}}
\def\ssb#1{_{\lower.25ex\hbox{$\scriptscriptstyle#1$}}}
\def\sbb#1{_{\lower.4ex\hbox{$\scriptstyle#1$}}}

\def\to{\rightarrow}
\def\mh{\ifmmode m\sbl H \else $m\sbl H$\fi}
\def\mch{\ifmmode m_{H^\pm} \else $m_{H^\pm}$\fi}
\def\mt{\ifmmode m_t\else $m_t$\fi}
\def\mc{\ifmmode m_c\else $m_c$\fi}
\def\mz{\ifmmode M_Z\else $M_Z$\fi}
\def\mw{\ifmmode M_W\else $M_W$\fi}
\def\mws{\ifmmode M_W^2 \else $M_W^2$\fi}
\def\mhs{\ifmmode m_H^2 \else $m_H^2$\fi}   
\def\mzs{\ifmmode M_Z^2 \else $M_Z^2$\fi}
\def\mts{\ifmmode m_t^2 \else $m_t^2$\fi}
\def\mcs{\ifmmode m_c^2 \else $m_c^2$\fi}
\def\mchs{\ifmmode m_{H^\pm}^2 \else $m_{H^\pm}^2$\fi}
\def\ztwo{\ifmmode Z_2\else $Z_2$\fi}
\def\zone{\ifmmode Z_1\else $Z_1$\fi}
\def\mtwo{\ifmmode M_2\else $M_2$\fi}
\def\mone{\ifmmode M_1\else $M_1$\fi}
\def\tb{\ifmmode \tan\beta \else $\tan\beta$\fi}
\def\xw{\ifmmode x\sub w\else $x\sub w$\fi}
\def\ch{\ifmmode H^\pm \else $H^\pm$\fi}
\def\lum{\ifmmode {\cal L}\else ${\cal L}$\fi}
\def\inpb{\ifmmode {\rm pb}^{-1}\else ${\rm pb}^{-1}$\fi}
\def\infb{\ifmmode {\rm fb}^{-1}\else ${\rm fb}^{-1}$\fi}
\def\epem{\ifmmode e^+e^-\else $e^+e^-$\fi}
\def\ppb{\ifmmode \bar pp\else $\bar pp$\fi}

\def\bsg{\ifmmode b\rightarrow s\gamma \else $b\rightarrow s\gamma$\fi}
 
\newskip\zatskip \zatskip=0pt plus0pt minus0pt
\def\matth{\mathsurround=0pt}

\def\atversim#1#2{\lower0.7ex\vbox{\baselineskip\zatskip\lineskip\zatskip
  \lineskiplimit 0pt\ialign{$\matth#1\hfil##\hfil$\crcr#2\crcr\sim\crcr}}}

\begin{document}
\rightline{\vbox{\halign{&#\hfil\cr
&SLAC-PUB-7250\cr
&August 1996\cr}}}
\vspace{0.8in} 
\Title{Determination of the $Z'$ Mass and Couplings Below Threshold at the NLC}
\bigskip
\Author{Thomas G. Rizzo\doeack}
\Address{\SLAC}
\bigskip
\begin{Abstract}
We investigate the capability of the NLC to indirectly determine both the 
mass as well as the couplings to leptons and $b$-quarks of a new neutral gauge 
boson below direct production threshold. By using data collected at several 
different values of the collider center of mass energy, we demonstrate how 
this can be done in an anonymous and model-independent manner. The procedure 
can be easily extended to the top and charm quark couplings.
\end{Abstract}
\bigskip
\vskip1.0in
\begin{center}
To appear in the {\it Proceedings of the 28th International 
Conference on High Energy Physics}, Warsaw, Poland, 25-31 July 1996
\end{center}
%
\bigskip
\def\thefootnote{\fnsymbol{footnote}}
\setcounter{footnote}{0}
\newpage
\section{Introduction}

A new neutral gauge boson, $Z'$, is the most well-studied of all exotic 
particles and is the hallmark signature for extensions of the Standard 
Model(SM) gauge group. 
If such a particle is found at future colliders the next step will be to 
ascertain its couplings to fermions. At hadron colliders, 
a rather long list of observables has been proposed over the years 
to probe these couplings--each with its own set of drawbacks 
and limitations~{\cite {rev}}. It has been shown, 
at least within the context of $E_6$-inspired 
models, that the LHC($\sqrt s=14$ TeV, $100fb^{-1}$) should be able 
to extract useful 
information on all of the $Z'$ couplings for $M_{Z'}$ below $\simeq 1-1.5$ 
TeV. (N.B., that this analysis included only statistical errors and should be 
redone.) At the NLC, when $\sqrt s < M_{Z'}$, a $Z'$ manifests itself only 
indirectly as deviations in observables, \eg, cross sections and asymmetries 
from their SM expectations. Fortunately the list 
of useful precision measurements that can be performed at the NLC is 
reasonably long and the expected large beam polarization($P$)  
plays a very important, perhaps crucial, role. In the past, analyses of the 
ability of 
the NLC to extract $Z'$ coupling information in this situation have taken 
for granted that the 
value of $M_{Z'}$ is already known from elsewhere, \eg, the LHC~{\cite {rev}}. 
Here we address the more complex issue of whether it is possible for the NLC to 
obtain information on the couplings of the $Z'$ if the mass were 
for some reason {\it a priori} unknown. In this case we would not only want to 
determine all fermionic couplings but in addition the $Z'$ mass as well.

If the $Z'$ mass were unknown it would appear that the traditional NLC $Z'$ 
coupling analyses would become problematic. Given 
data at a fixed value of $\sqrt s$ which shows deviations from the SM due to 
a $Z'$, one 
would not be able to {\it simultaneously} extract the value of $M_{Z'}$ as 
well as the corresponding couplings. The reason is clear: to leading order in 
$s/M_{Z'}^2$, rescaling 
all of the couplings and the value of $M_{Z'}$ by a common factor 
would leave the observed deviations from the SM invariant. In this 
approximation, the $Z'$ exchange appears only as a contact interaction. 
Thus as long as $\sqrt s < M_{Z'}$, the only potential solution to this 
problem lies in obtaining data on deviations in observables from the 
SM at {\it several}, distinct $\sqrt s$ values and combining them into a 
single fit. Here we report on the first analysis 
of this kind, focussing on observables involving only leptons and/or  
$b$-quarks. In performing such an analysis there are many questions that one 
can raise: how many $\sqrt s$ values are 
needed? How do we distribute the luminosity($\cal L$) to optimize the 
results? Can such an analysis be done while maintaining model-independence? 
In this {\it initial} study we begin to address these and some related 
questions. 

\section{Analysis}

In order to proceed with this benchmark study, we will make a number of 
simplifying assumptions and parameter choices. These can be modified at a 
later stage to 
see how they influence our results. In this analysis we consider the following 
ten observables: $\sigma_{\ell,b}$, $A_{FB}^{\ell,b}$, $A_{LR}^{\ell,b}$, 
$A_{pol}^{FB}(\ell,b)$, $<P_\tau>$, and $P_\tau^{FB}$. Other inputs and 
assumptions are as follows:
\vspace*{0.2cm}
\begin{tabbing}
   Anomalous Trilinear Gauge Couplings Fun \= 8.Fun and games with Extended 
Technicolor models \kill
   ~~~~~e,$\mu$,$\tau$ universality \> ISR with $\sqrt {s'}/\sqrt {s} >0.7$ \\
   ~~~~~$P=90\%$, $\delta P/P=0.3\%$  \> $\delta {\cal L}/ {\cal L}=0.25\%$ \\
   ~~~~~$\epsilon_b=50\%$, $\Pi_b=100\%$  \> $|\theta|>10^\circ$ \\
   ~~~~~$\epsilon_{e,\mu,\tau}(\sigma)=100\%$, $\epsilon_\tau(P_\tau)=50\%$ 
\> $\delta A/A=0.3\%$ \\
   ~~~~~Neglect $t$-channel terms in $e^+e^-\to e^+e^-$ \>  \\
\end{tabbing}
Of special note on this list are ($i$) a $b$-tagging 
efficiency($\epsilon_b$) of $50\%$ for a purity($\Pi_b$) of 100$\%$, ($ii$) 
the efficiency for identifying all leptons is assumed to be 100$\%$, although 
only $50\%$ of $\tau$ decays are assumed to be polarization analyzed, ($iii$) 
a $10^\circ$ angle cut has been applied to all final state fermions, and 
($iv$) a strong cut to  events with an excess of initial state 
radiation(ISR) has also been made. ($v$) A small systematic error associated 
with the various asymmetry measurements has also been included. 
In addition to the above, final state QED as well as QCD corrections are 
included, the 
$b$-quark and $\tau$ masses have been neglected, and the possibility of 
$Z-Z'$ mixing has been ignored. Since 
our results will generally be statistics limited, the role played by the 
systematic uncertainties associated with the parameter choices above will 
generally be rather minimal. 

To insure model-independence, the values of the $Z'$ couplings, \ie, 
$(v,a)_{\ell,b}$, as well as $M_{Z'}$, are chosen {\it randomly} and 
{\it anonymously} from 
rather large ranges representative of a number of extended gauge models. 
Monte Carlo data representing the above observables 
is then generated for several different values of $\sqrt s$. At this point, the 
values of the mass and couplings are not `known'  
{\it a priori}, but will later be compared with what is extracted 
from the Monte Carlo generated event sample. Following this approach 
there is no particular relationship between any of the couplings and 
there is no dependence upon any particular $Z'$ model. (We normalize our 
couplings so that 
for the SM $Z$, $a_\ell=-1/2$.) Performing this analysis for a wide range of 
possible mass and coupling choices then shows the power as well as the 
limitations of this technique. 

To get an understanding for how this procedure works in general we will make 
two case studies for the $Z'$ mass and couplings, labelled here by I and II. 
There is nothing special about these two choices and several other parameter 
sets have been analyzed in comparable detail to show that the results that 
we display below are rather typical. To begin 
we generate Monte Carlo data at $\sqrt {s}=$0.5, 0.75 and 1 TeV 
with associated integrated luminosities of 70, 100, and 150 
$fb^{-1}$, respectively, and subsequently 
determine the 5-dimensional $95\%$ CL allowed region for the mass and 
couplings from a simultaneous fit using the assumptions listed above. This 
5-dimensional region is then 
projected into a series of 2-dimensional plots which we can examine in detail. 
Figs. 1 and 2 show the results of our analysis for these two case studies 
compared 
with the expectations of a number of well-known $Z'$ models~{\cite {rev}}. 
Several things are immediately apparent-the most obvious being that two 
distinct allowed regions are obtained from the fit in both cases. (The input 
values are seen to lie nicely inside one of them.) This two-fold ambiguity 
results from our inability to make a  
determination of the overall sign of {\it one} of the couplings, \eg, 
$a_\ell$. If the sign 
of  $a_\ell$ were known, only a single allowed region would appear in 
Figs. 1a-b and 2a-b and a unique coupling determination would be obtained. 
Note that this {\it same} sign ambiguity arises in SLD/LEP data for the 
SM $Z$ and is only removed through the examination of low-energy neutrino 
scattering. Secondly, we see that the leptonic couplings 
are somewhat better determined than are those of the $b$-quark, which is 
due to the fact that the leptonic 
observables involve only leptonic couplings, while 
those for $b$-quarks involve both 
types. In addition, there is more statistical power available in the lepton 
channels due to the assumption of universality and the 
leptonic results employ two additional observables related to $\tau$ 
polarization. Thirdly, we see from 
Figs. 1a-b the importance in obtaining coupling information for a number of 
different fermion species. If only the Fig. 1a results were available, one 
might draw the hasty conclusion that an $E_6$-type $Z'$ had been found. Fig. 1b 
clearly shows us that this is not the case. Evidently {\it neither} $Z'$ 
corresponds to any well-known model. Lastly, as promised, the $Z'$ mass is 
determined in both cases, although with somewhat smaller uncertainties in 
case II. We remind the reader that there is nothing special about these two 
particular cases. 

\vspace*{-0.5cm}
\nn
\begin{figure}[htbp]
\centerline{
\psfig{figure=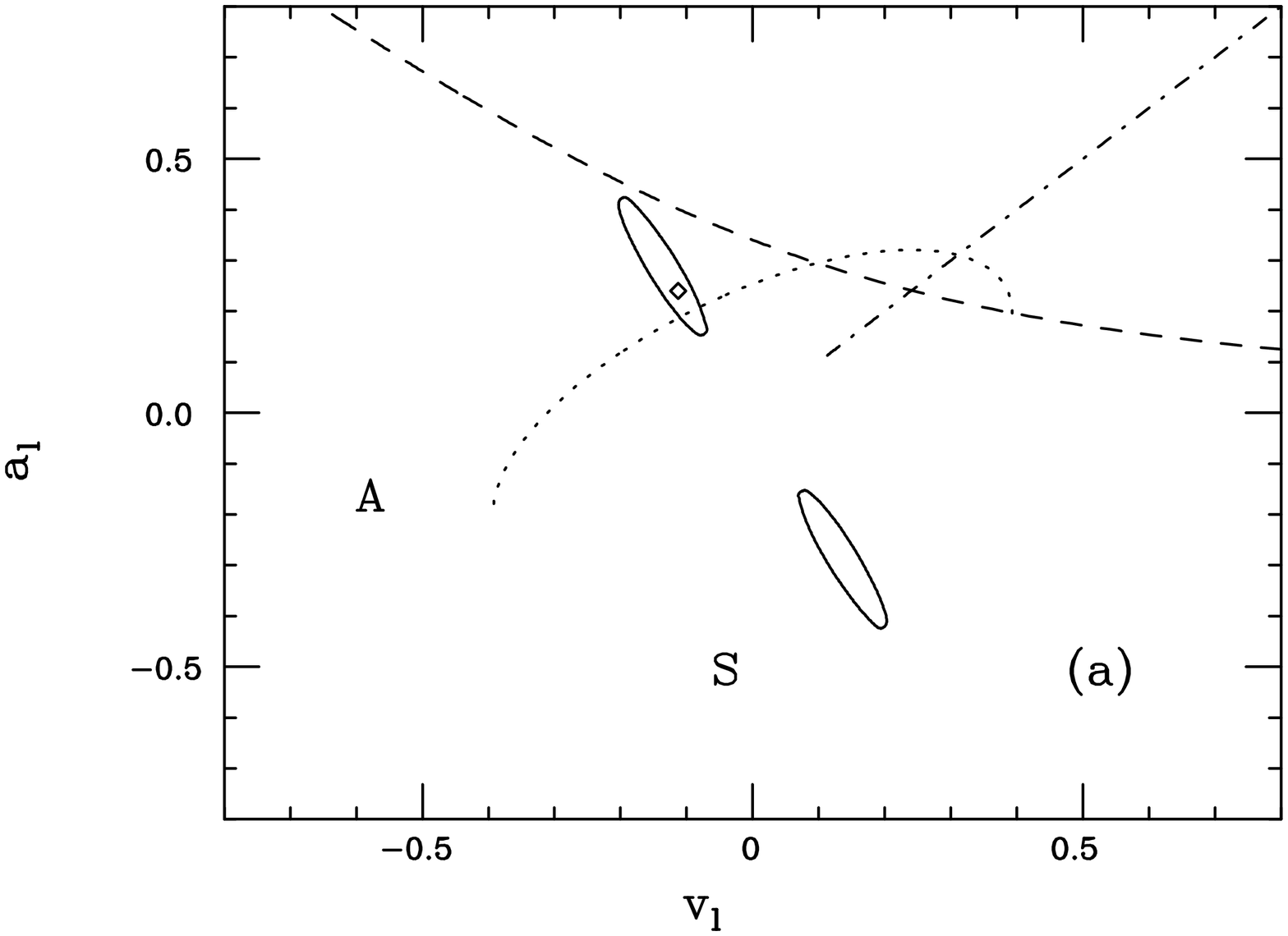,height=9.1cm,width=9.1cm,angle=-90}
\hspace*{-5mm}
\psfig{figure=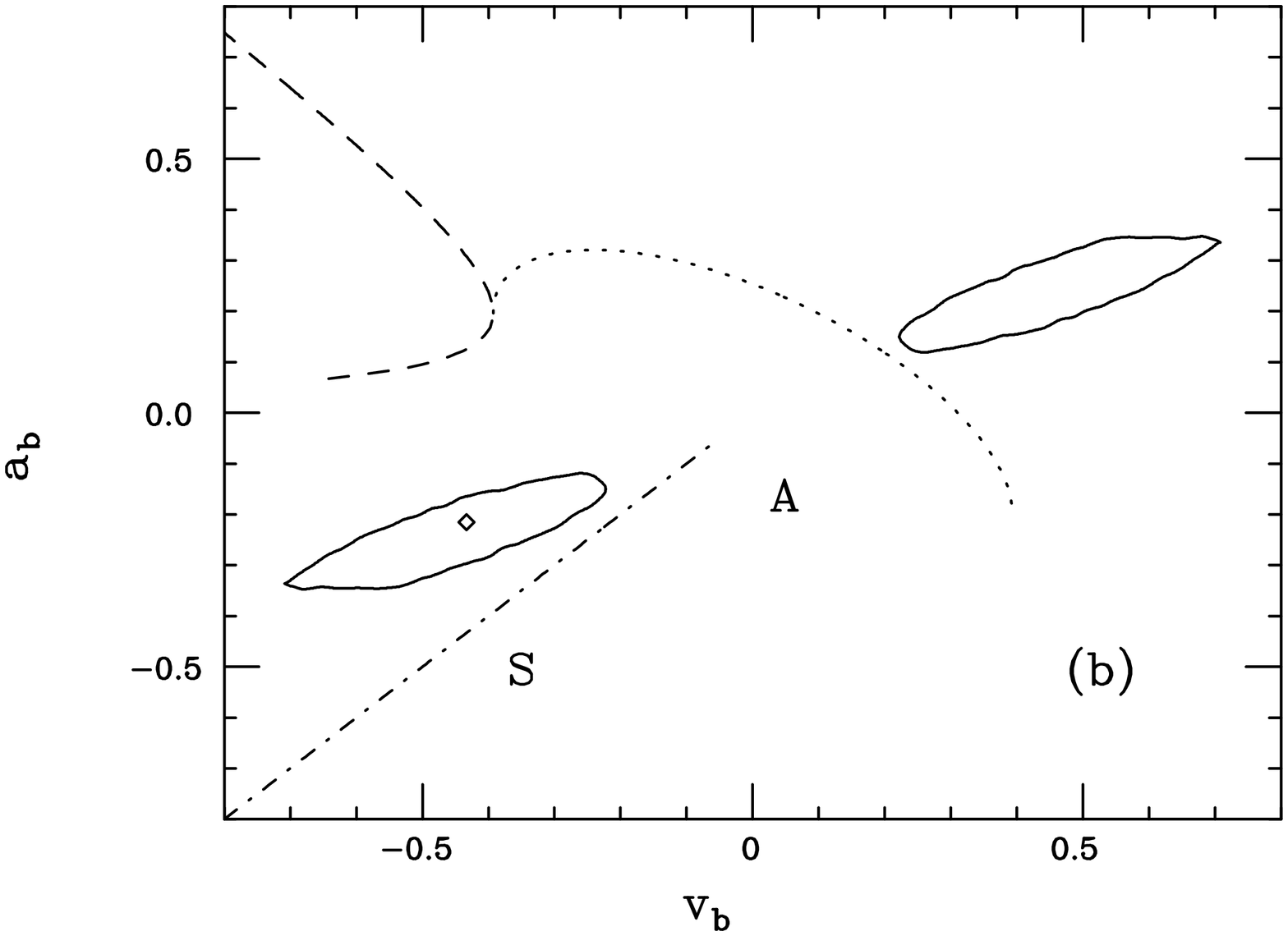,height=9.1cm,width=9.1cm,angle=-90}}
\vspace*{-0.75cm}
\centerline{
\psfig{figure=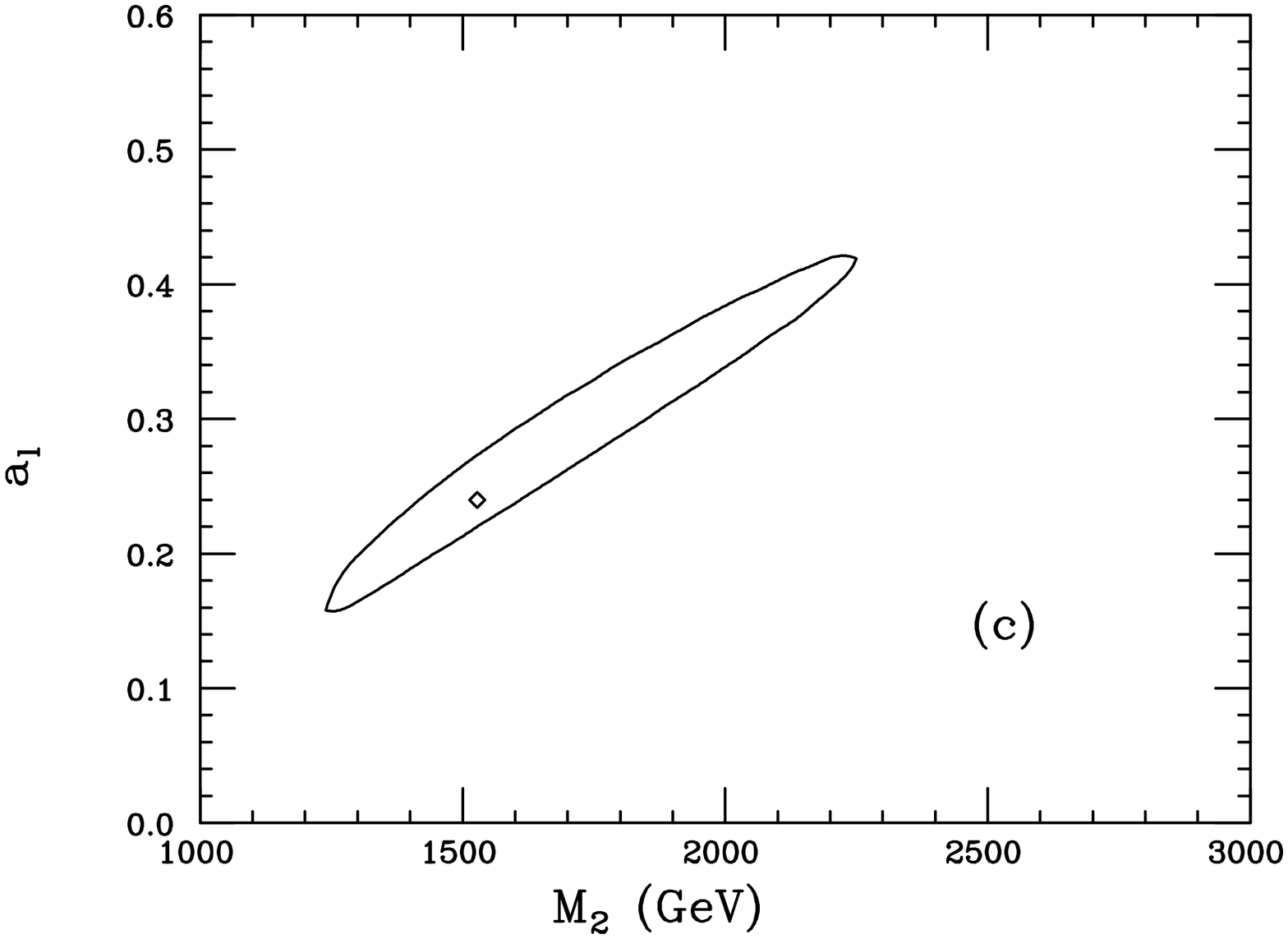,height=9.1cm,width=9.1cm,angle=-90}}
\vspace*{-0.3cm}
\caption{\small $95\%$ CL allowed regions for the extracted values of the 
(a) lepton and (b) $b$-quark couplings 
for the $Z'$ of case I compared with the predictions of the $E_6$ 
model(dotted), the Left-Right Model(dashed), and the Un-unified 
Model(dash-dot), 
as well as the Sequential SM and Alternative LR Models(labeled by `S' and `A', 
respectively.) (c) Extracted $Z'$ mass; only the $a_\ell >0$ branch is shown. 
In all cases the diamond represents the corresponding input values.}
\end{figure}
\vspace*{-0.5cm}
\nn
\begin{figure}[htbp]
\centerline{
\psfig{figure=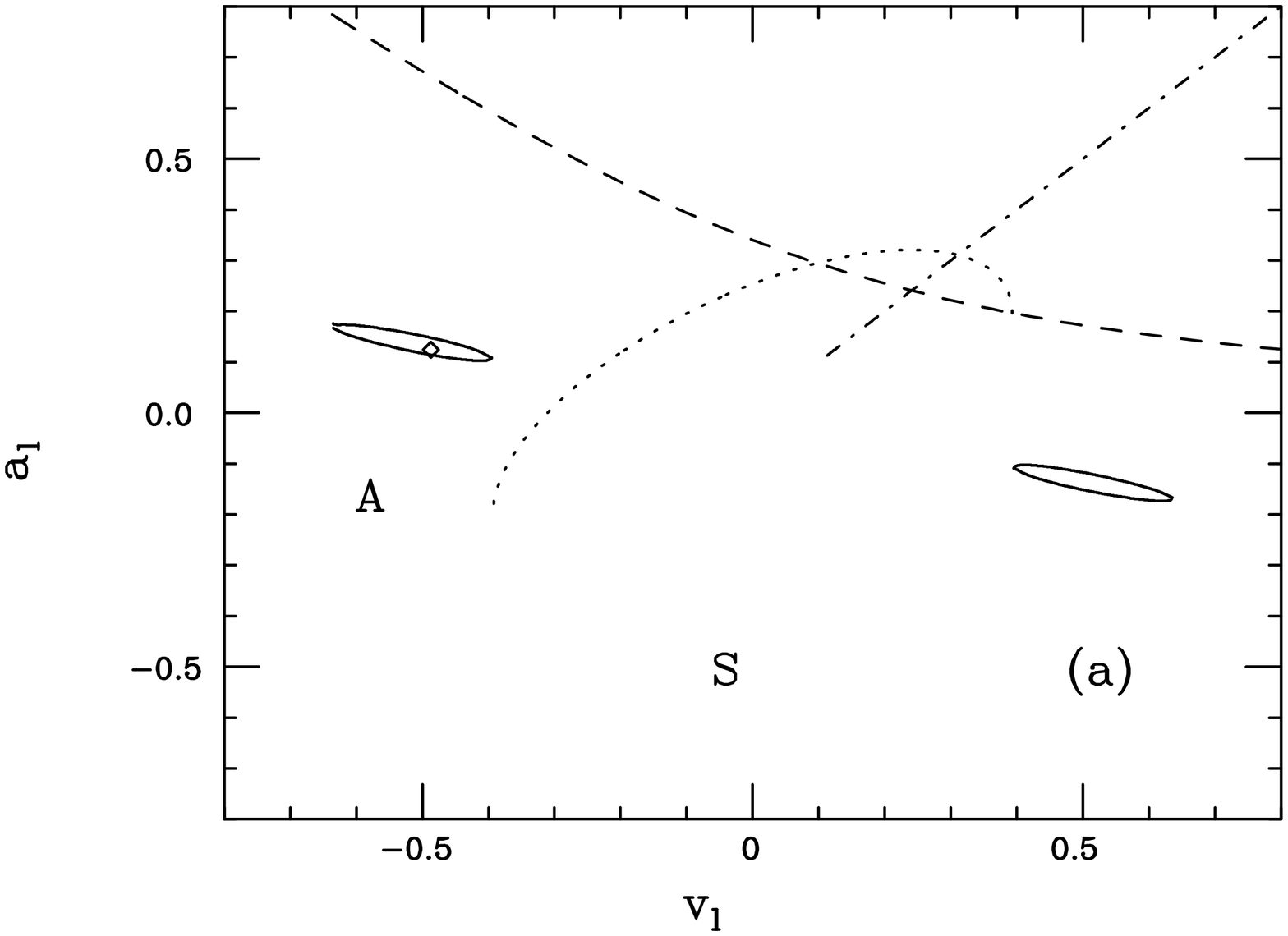,height=9.1cm,width=9.1cm,angle=-90}
\hspace*{-5mm}
\psfig{figure=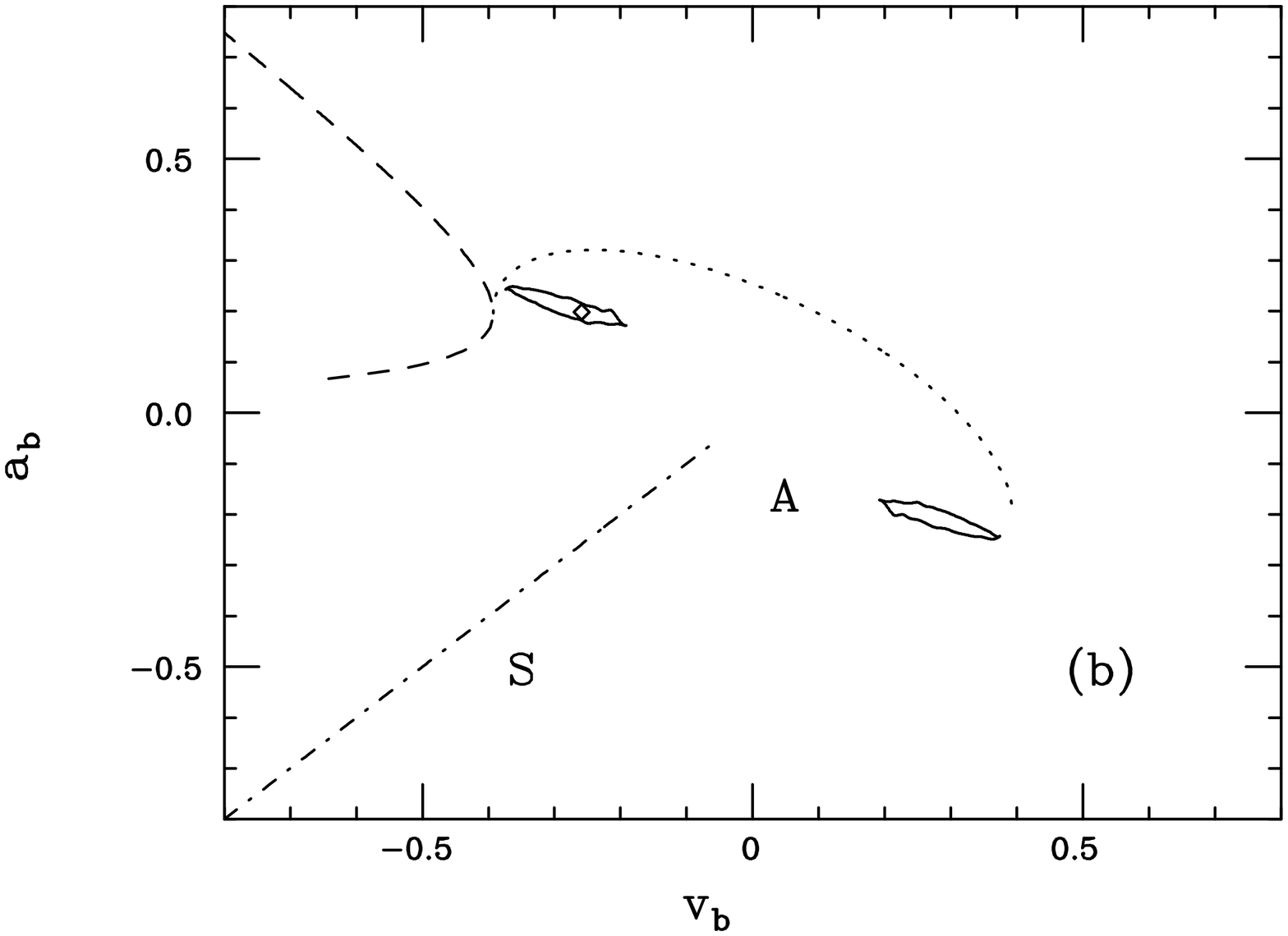,height=9.1cm,width=9.1cm,angle=-90}}
\vspace*{-0.75cm}
\centerline{
\psfig{figure=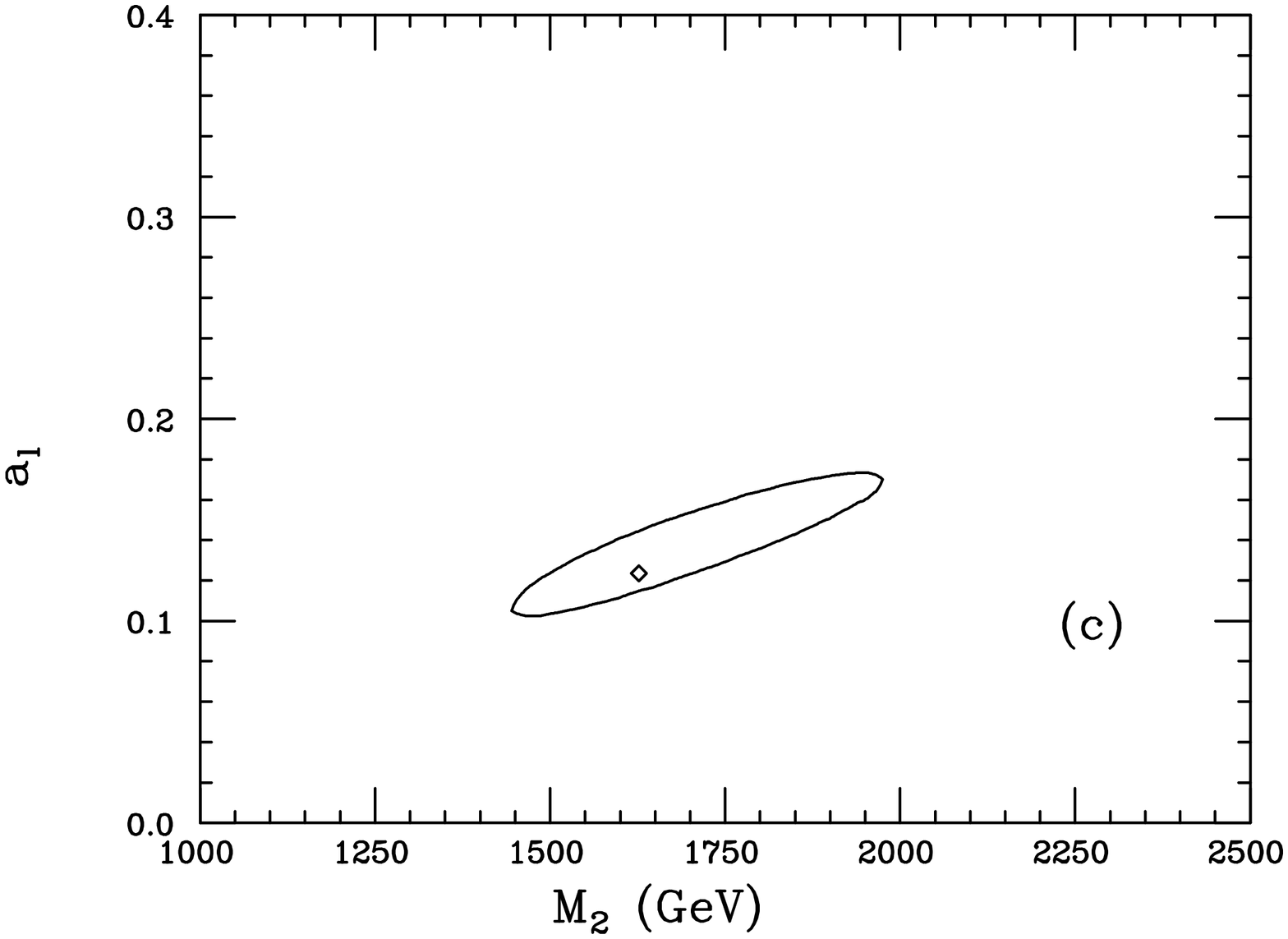,height=9.1cm,width=9.1cm,angle=-90}}
\vspace*{-0.3cm}
\caption{\small Same as Fig. 1 but for a different choice of $Z'$ mass and 
couplings referred to as case II in the text.}
\end{figure}

Of course, the clever reader must now be asking the question `why use 3 
different 
values of $\sqrt s$, why not 2 or 5?' This is a very important issue which we 
can only begin to address here. Let us return to the mass and couplings of 
case I and generate Monte Carlo `data' for 
only $\sqrt s$=0.5 and 1 TeV with $\cal L$= 100 and 220 $fb^{-1}$, 
respectively, thus keeping the total $\cal L$ the {\it same} as in the 
discussion above. Repeating our analysis we then 
arrive at the `2-point' fit as shown in Fig. 3a; unlike Fig. 1a, the 
allowed region does not 
close and extends outward to ever larger values of 
$v_\ell,a_\ell$. The corresponding 
$Z'$ mass contour also does not close, again extending outwards to ever larger 
values. We realize immediately that this is just what happens when data at 
only a single $\sqrt s$ is available. For our fixed $\cal L$, distributed as we 
have done, we see that there is not enough of a lever arm to simultaneously 
disentangle the $Z'$ mass and 
couplings. Of course the reverse situation can also be just as bad. We   
now generate Monte Carlo `data' for the case I mass and couplings in 100 GeV 
steps in $\sqrt s$ over the 0.5 to 1 TeV interval with the same total 
$\cal L$ as above but now distributed as 30, 30, 50, 50, 60, and 100 $fb^{-1}$, 
respectively. We then arrive at the `6-point' fit shown in Fig. 3b 
which suffers 
a problem similar to Fig. 3a. What has happened now is that we have spread 
the fixed $\cal L$ too thinly over too many points for the 
analysis to work. This brief study indicates that a proper balance is 
required to simultaneously achieve the desired statistics as well as an 
effective lever arm to obtain the $Z'$ mass and couplings. It is important 
to remember that we 
have {\it not} demonstrated that the `2-point' fit will never work. We note 
only that it fails with our specific fixed luminosity distribution for the 
masses and couplings associated with cases I and II. It is possible that for 
`lucky' combinations of masses and couplings a 2-point fit will suffice. 
Clearly, more work is required to further address this issue.  

\vspace*{-0.5cm}
\nn
\begin{figure}[htbp]
\centerline{
\psfig{figure=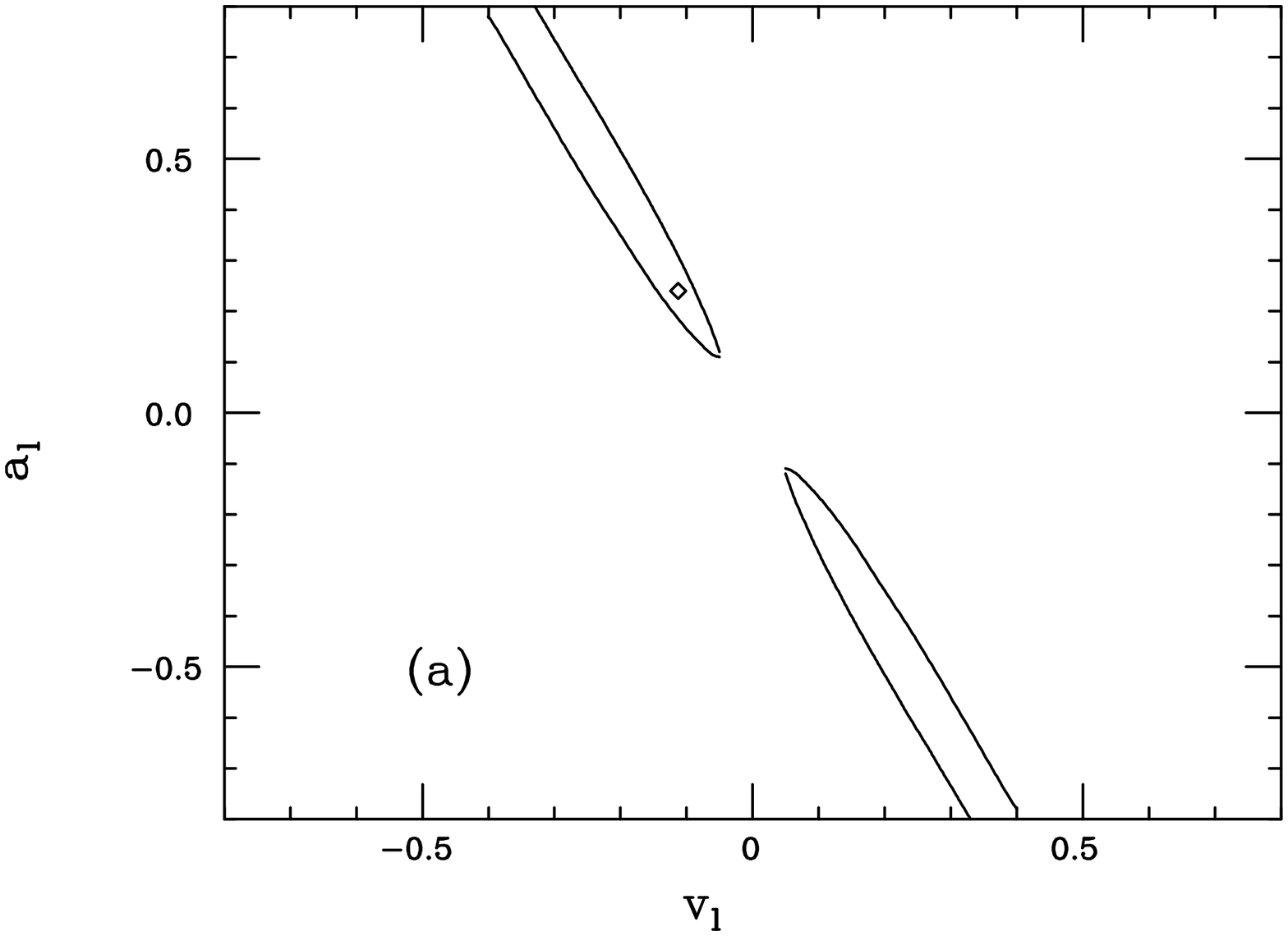,height=9.1cm,width=9.1cm,angle=-90}
\hspace*{-5mm}
\psfig{figure=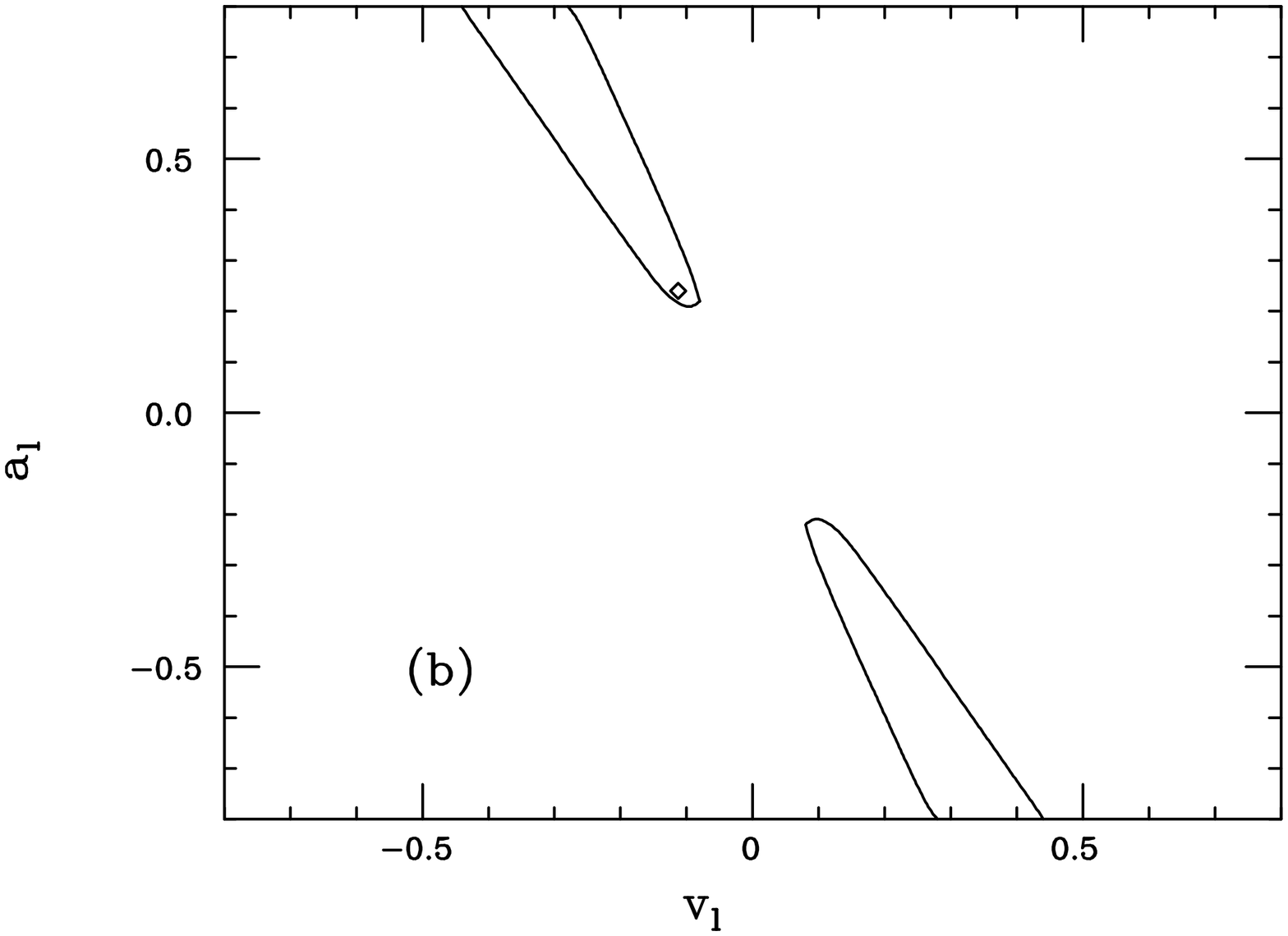,height=9.1cm,width=9.1cm,angle=-90}}
\vspace*{-0.3cm}
\caption{\small Failure of the method in case I when data is taken at 
(a) too few (`2-point' fit) or (b) too many (`6-point' fit) different 
center of mass energies for the same total integrated 
luminosity as in Figs. 1 and 2. The luminosities are distributed as discussed 
in the text.}
\end{figure}
\vspace*{0.4mm}

How do these results change if $M_{Z'}$ {\it were} known or if our input 
assumptions were modified? Let us return to case I and concentrate on the 
allowed 
coupling regions corresponding to a choice of negative values of 
$v_{\ell,b}$; these are 
expanded to the solid curves shown in Figs. 4a and 4c. The large dashed curve 
in Fig. 4a corresponds to a reduction of the polarization to $80\%$ with the 
same relative error as before. While the allowed region expands the 
degradation is not severe. If the $Z'$ mass were known, the `large'  
ellipses shrink to 
the small ovals in Fig. 4a; these are expanded in Fig. 4b. This is clearly a 
radical reduction in the size of the allowed region! We see that when the 
mass is known, varying the polarization or its uncertainty over a reasonable 
range has very little influence on the resulting size of the allowed 
regions. From Fig. 4c we see that while knowing the $Z'$ mass significantly 
reduces the size of the allowed region for the $b$ couplings, the impact is 
far less than in the leptonic case 
for the reasons discussed above. Figs. 5a and 5b show that case I is not 
special in that similar results are seen to hold for case II.

\vspace*{-0.5cm}
\nn
\begin{figure}[htbp]
\centerline{
\psfig{figure=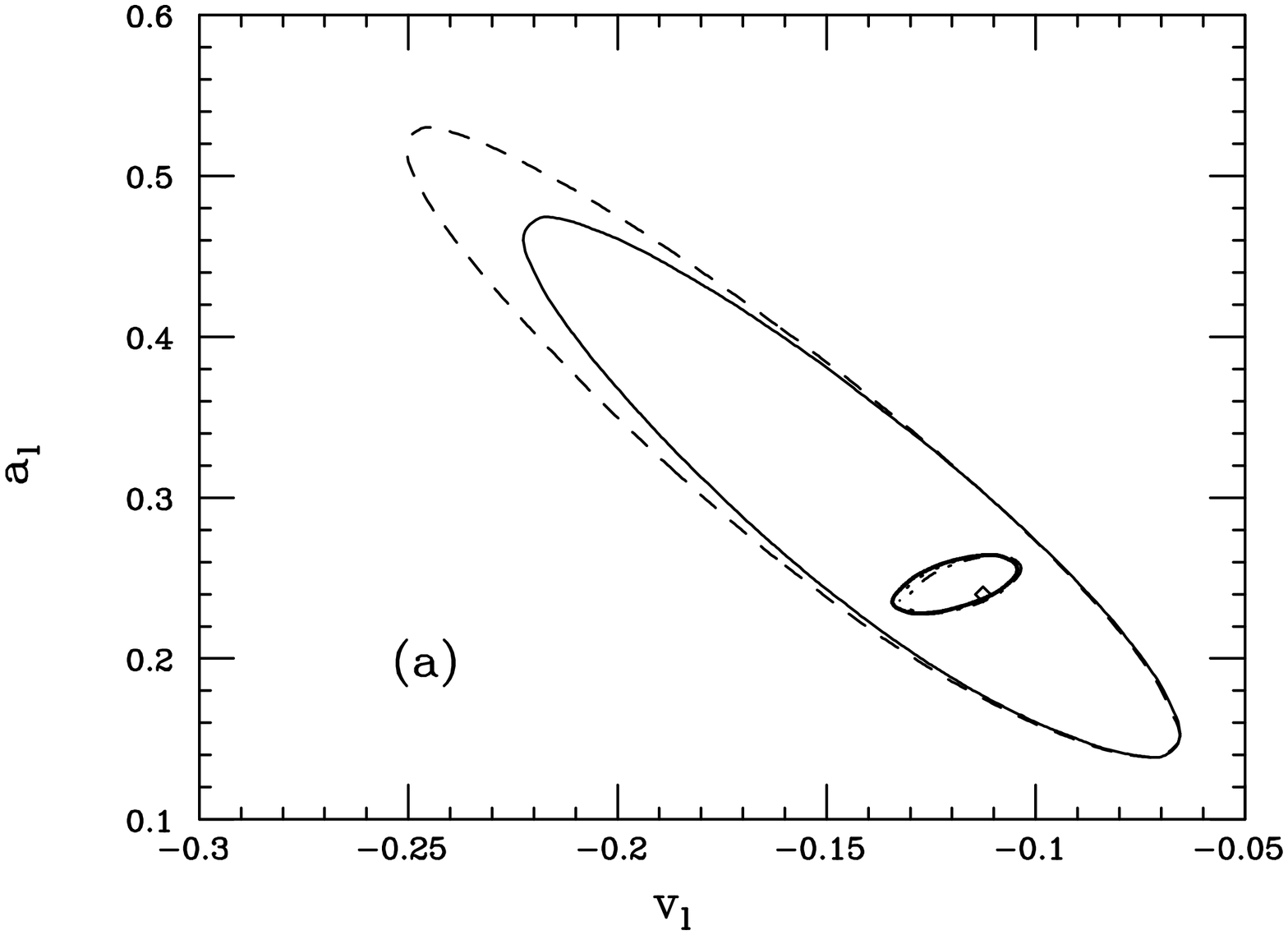,height=9.1cm,width=9.1cm,angle=-90}
\hspace*{-5mm}
\psfig{figure=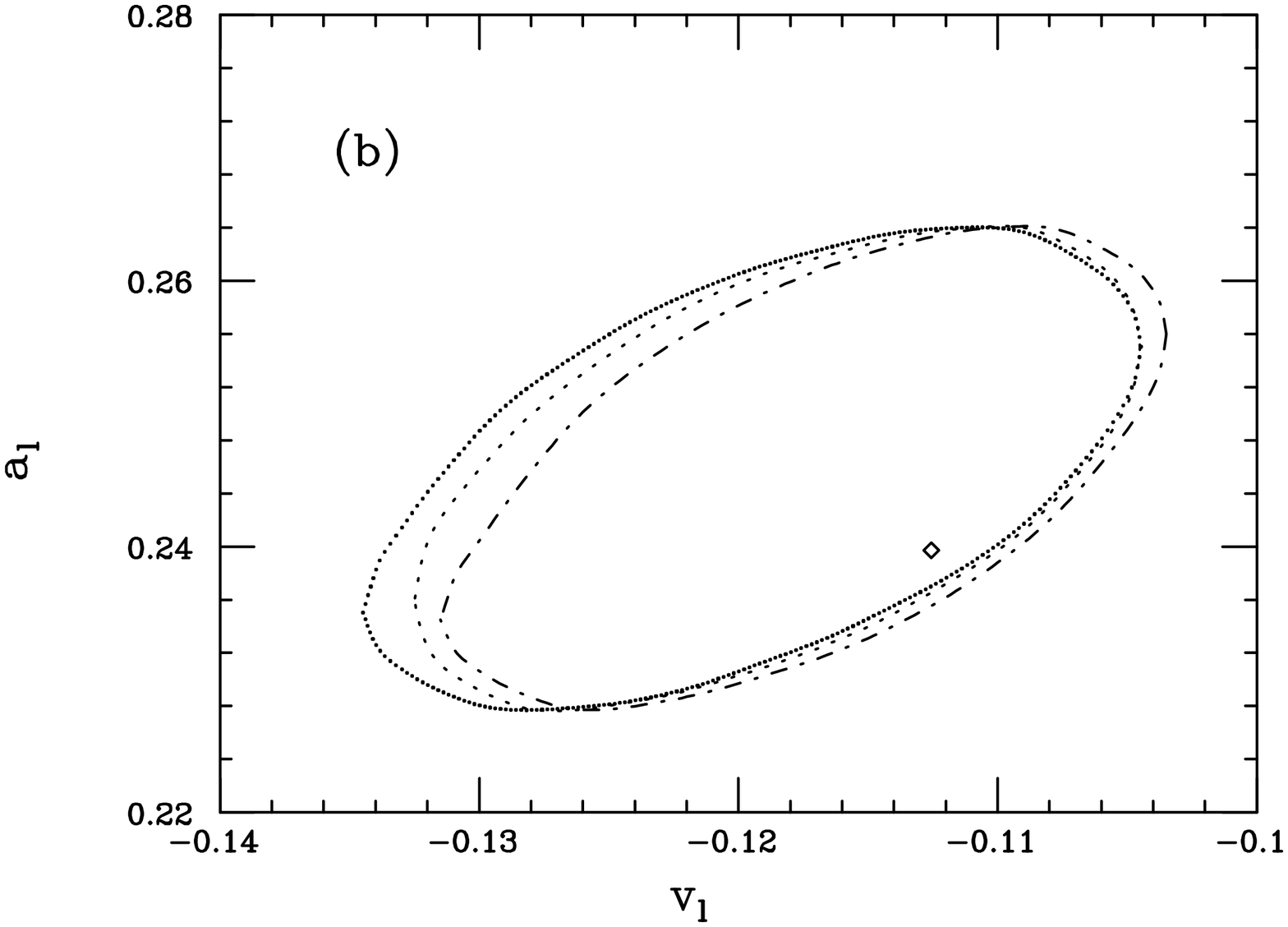,height=9.1cm,width=9.1cm,angle=-90}}
\vspace*{-0.75cm}
\centerline{
\psfig{figure=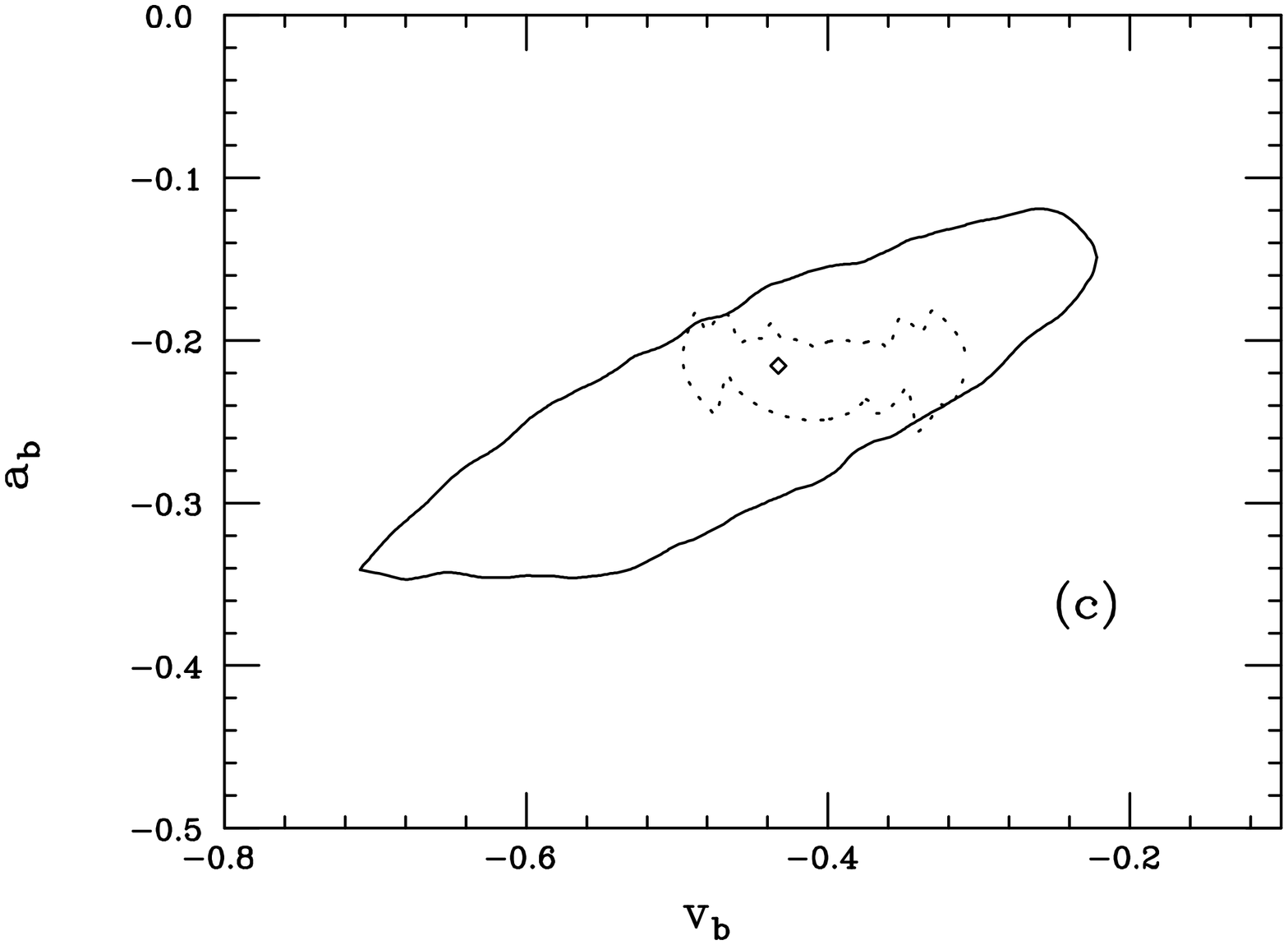,height=9.1cm,width=9.1cm,angle=-90}}
\vspace*{-0.3cm}
\caption{\small (a) Expanded lobe(solid) from Fig. 1a; the dashed curve shows 
the same result but for $P=80\%$. The smaller ovals, expanded in (b) apply 
when the $Z'$ mass is known. Here, in (b), $P=90(80)\%$ corresponds to the 
dash-dot(dotted) curve while the case of $P=90\%$ with $\delta P/P=5\%$ 
corresponds to the square-dotted curve. (c) Expanded lobe(solid) from Fig.1b; 
the dotted curve corresponds to the case when $M_{Z'}$ is known.} 
\end{figure}

What happens for larger $Z'$ masses or when data at larger values of $\sqrt s$ 
becomes available? Let us assume that the `data' from the above three center 
of mass 
energies is already existent, with the luminosities as given. We now imagine 
that the NLC increases its center of mass energy to $\sqrt s$= 1.5 TeV and 
collects an additional 
200 $fb^{-1}$ of integrated luminosity. Clearly for $Z'$ masses near or 
below 1.5 TeV our 
problems are solved since an on-shell $Z'$ can now be produced. Thus we 
shall concern ourselves with $Z'$ masses in excess of 2 TeV. 
Figs. 6a-d  show the result of extending our procedure--now using 4 
different $\sqrt s$ values, again for two distinct choices of the 
$Z'$ mass and 
couplings. These `4-point' results are a combined fit to the data at 
all four center of mass energies. 
(Only one of the allowed pair of ellipses resulting from the overall sign 
ambiguity is shown for simplicity.) 
Note that the $Z'$ input masses we have chosen are well in excess of 2 TeV 
where the LHC may provide only very minimal information on the fermion 
couplings~{\cite {rev}}. Clearly by using the additional data from a 
run at $\sqrt s$=1.5 TeV this technique can be extended to perform coupling 
extraction for $Z'$ masses in excess of 2.5 TeV. The maximum `reach' for the 
type of coupling analysis we have done is not yet known. It seems likely, 
based on these initial studies, that the extraction of interesting coupling 
information for $Z'$ masses in excess of 3 TeV seems possible for a reasonable 
range of parameters.

\vspace*{-0.5cm}
\nn
\begin{figure}[htbp]
\centerline{
\psfig{figure=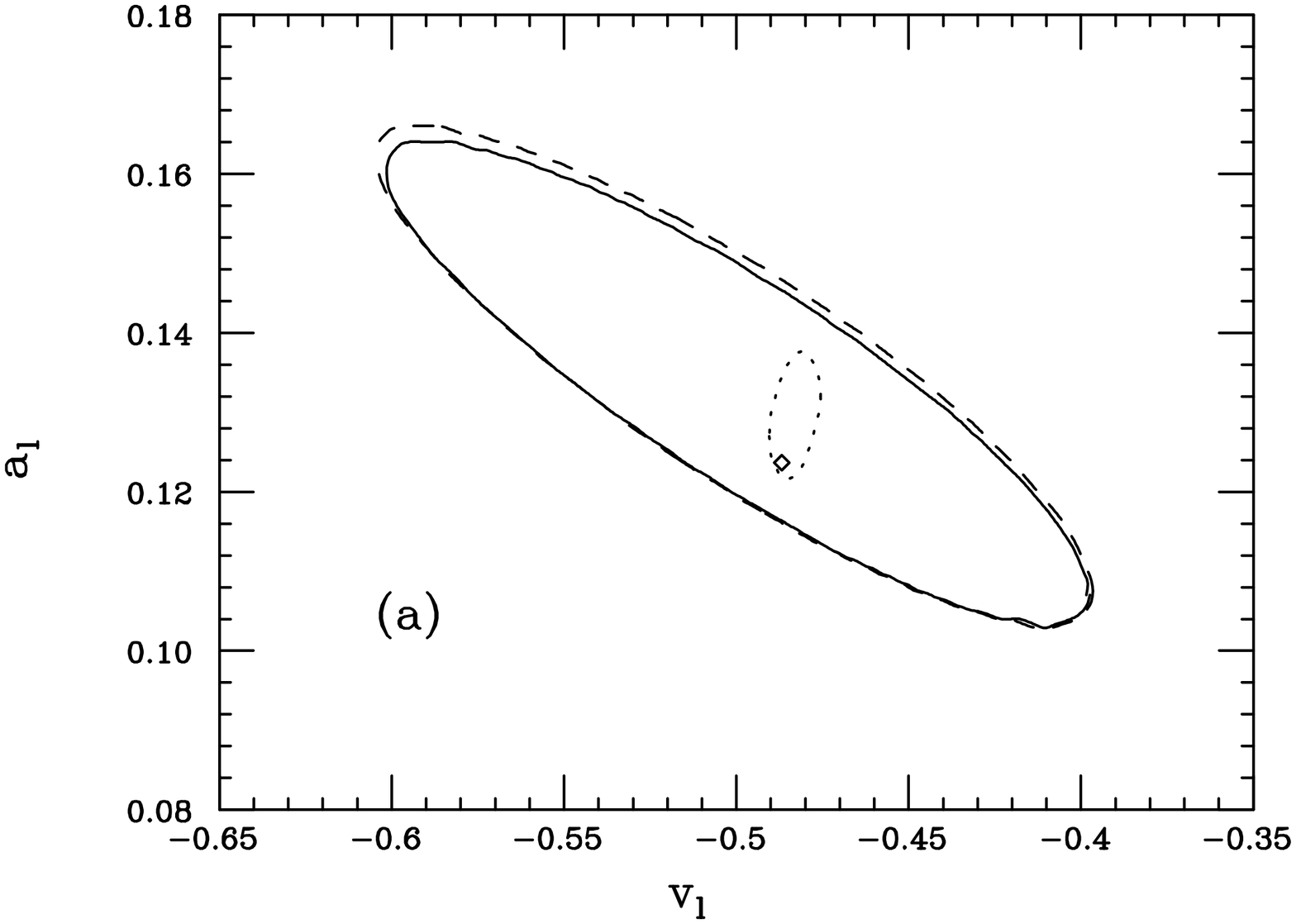,height=9.1cm,width=9.1cm,angle=-90}
\hspace*{-5mm}
\psfig{figure=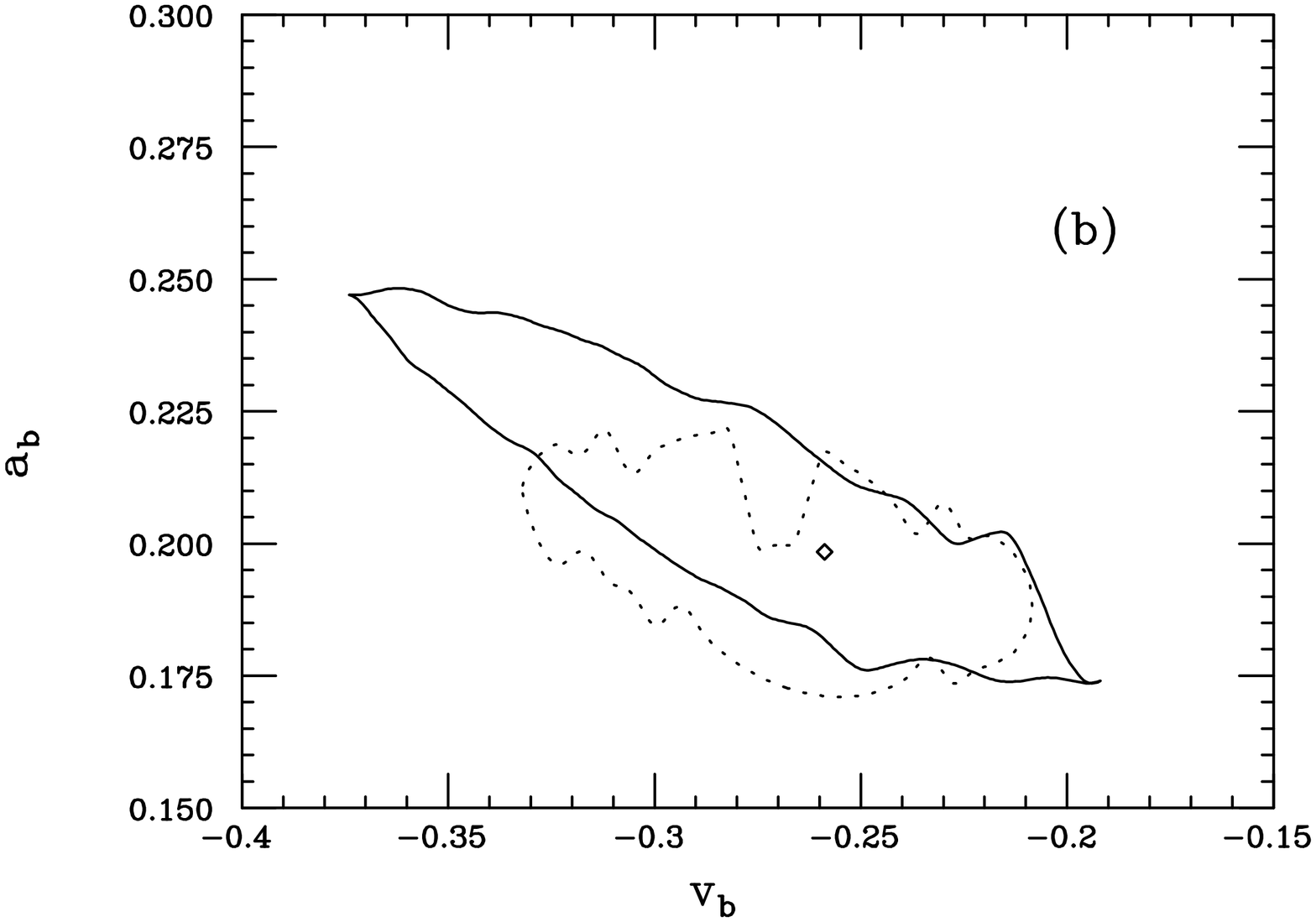,height=9.1cm,width=9.1cm,angle=-90}}
\vspace*{-0.3cm}
\caption{\small (a) Expanded lobe(solid) from Fig. 2a; the dashed curve shows 
the same result but for $P=80\%$. The smaller dotted oval, applies  
when the $Z'$ mass is known and $P=90\%$. (b) Expanded lobe(solid) from 
Fig. 2b; 
the dotted curve corresponds to the case when $M_{Z'}$ is known. }
\end{figure}
\vspace*{0.4mm}

\section{Outlook and Conclusions}

In this paper we have shown that it is possible for the NLC to extract 
information on the $Z'$ couplings to leptons and $b$-quarks even when the $Z'$ 
mass is not {\it a priori} known. The critical step for the success of the 
analysis was to combine the data available from measurements performed 
at several different center of mass energies. For reasonable luminosities 
the specific results we have obtained suggest, but do not prove, that data 
sets at at least 3 different energies are necessary for the procedure to be 
successful.

\vspace*{-0.5cm}
\nn
\begin{figure}[htbp]
\centerline{
\psfig{figure=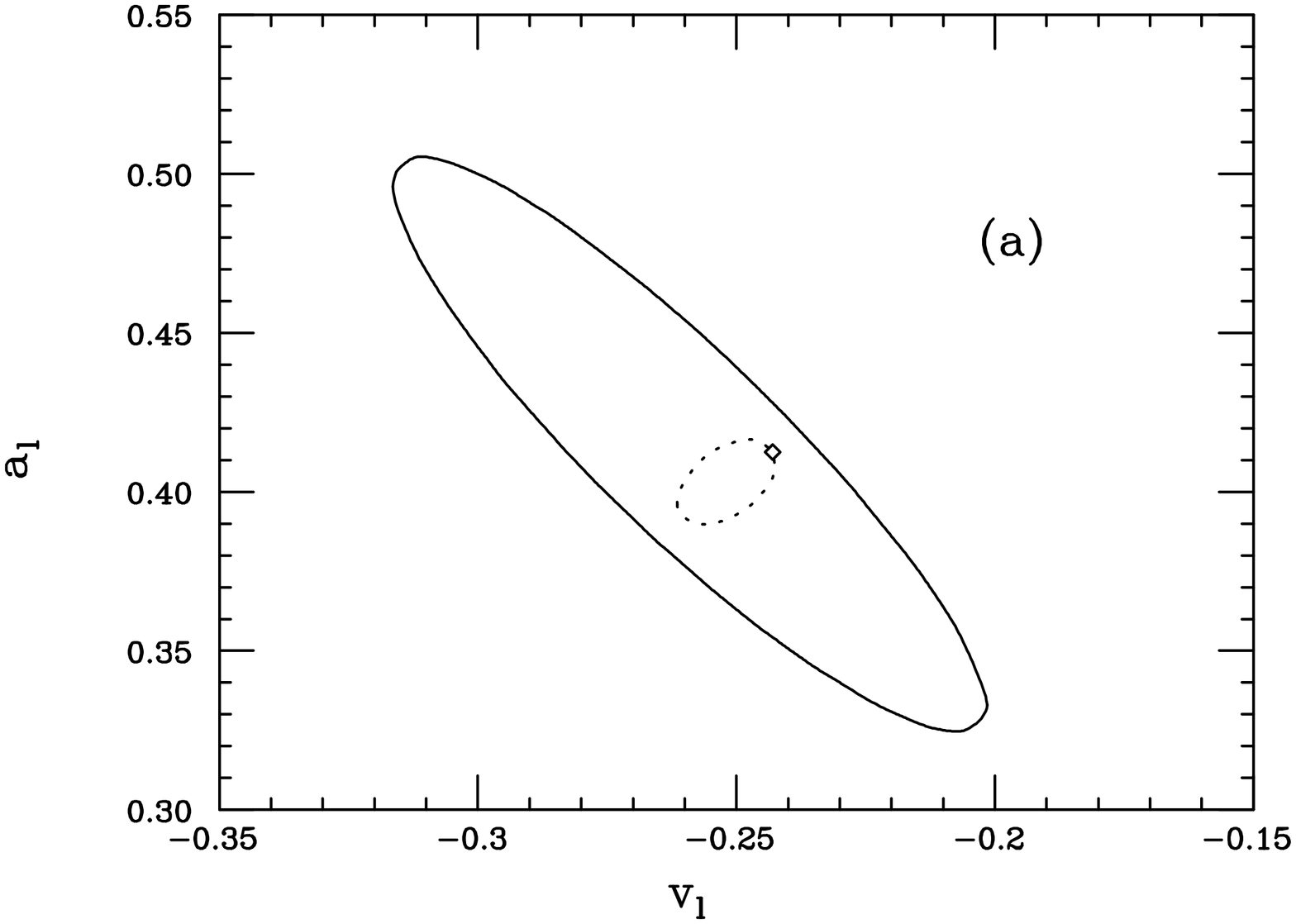,height=9.1cm,width=9.1cm,angle=-90}
\hspace*{-5mm}
\psfig{figure=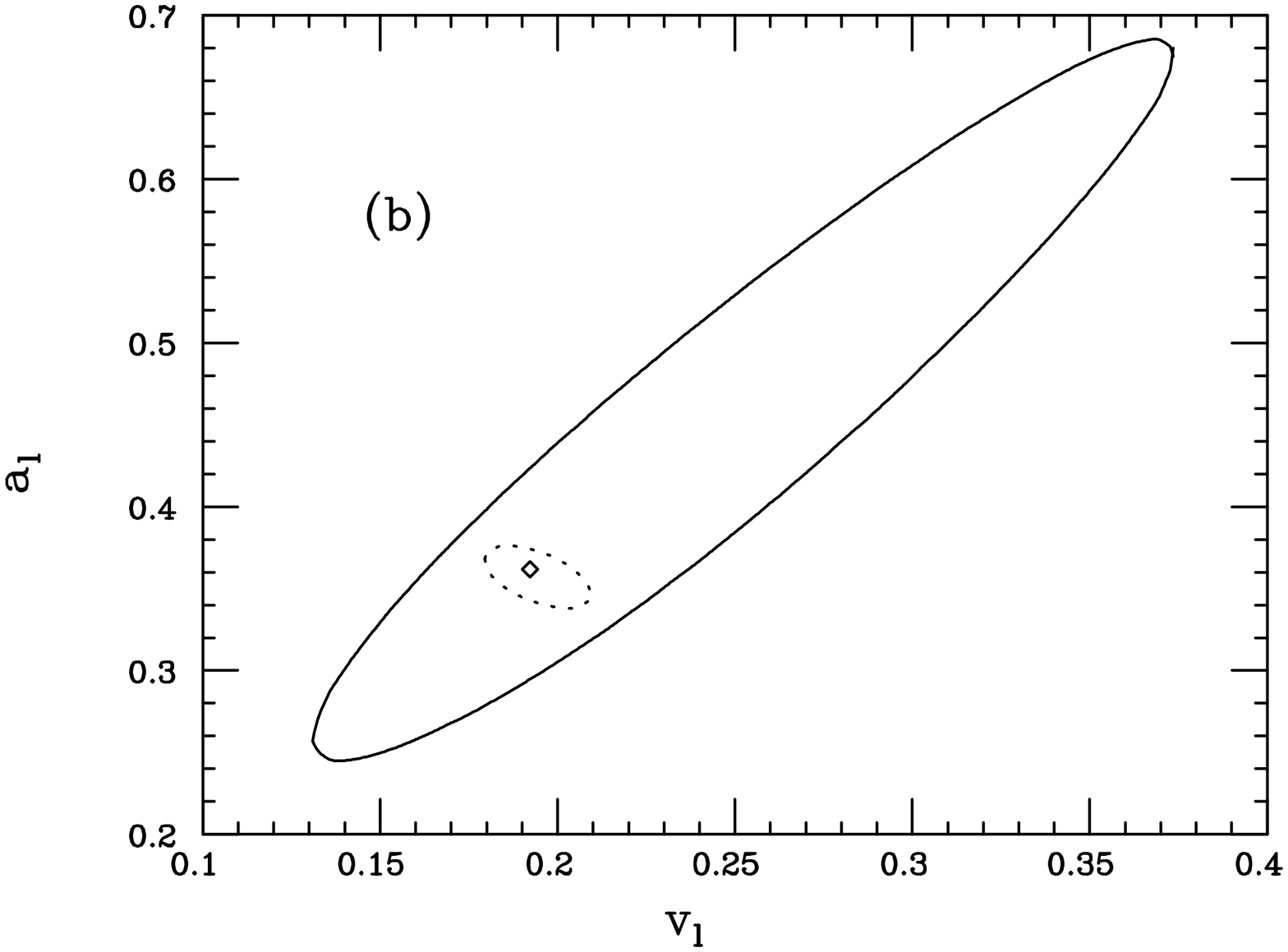,height=9.1cm,width=9.1cm,angle=-90}}
\vspace*{0.1cm}
\centerline{
\psfig{figure=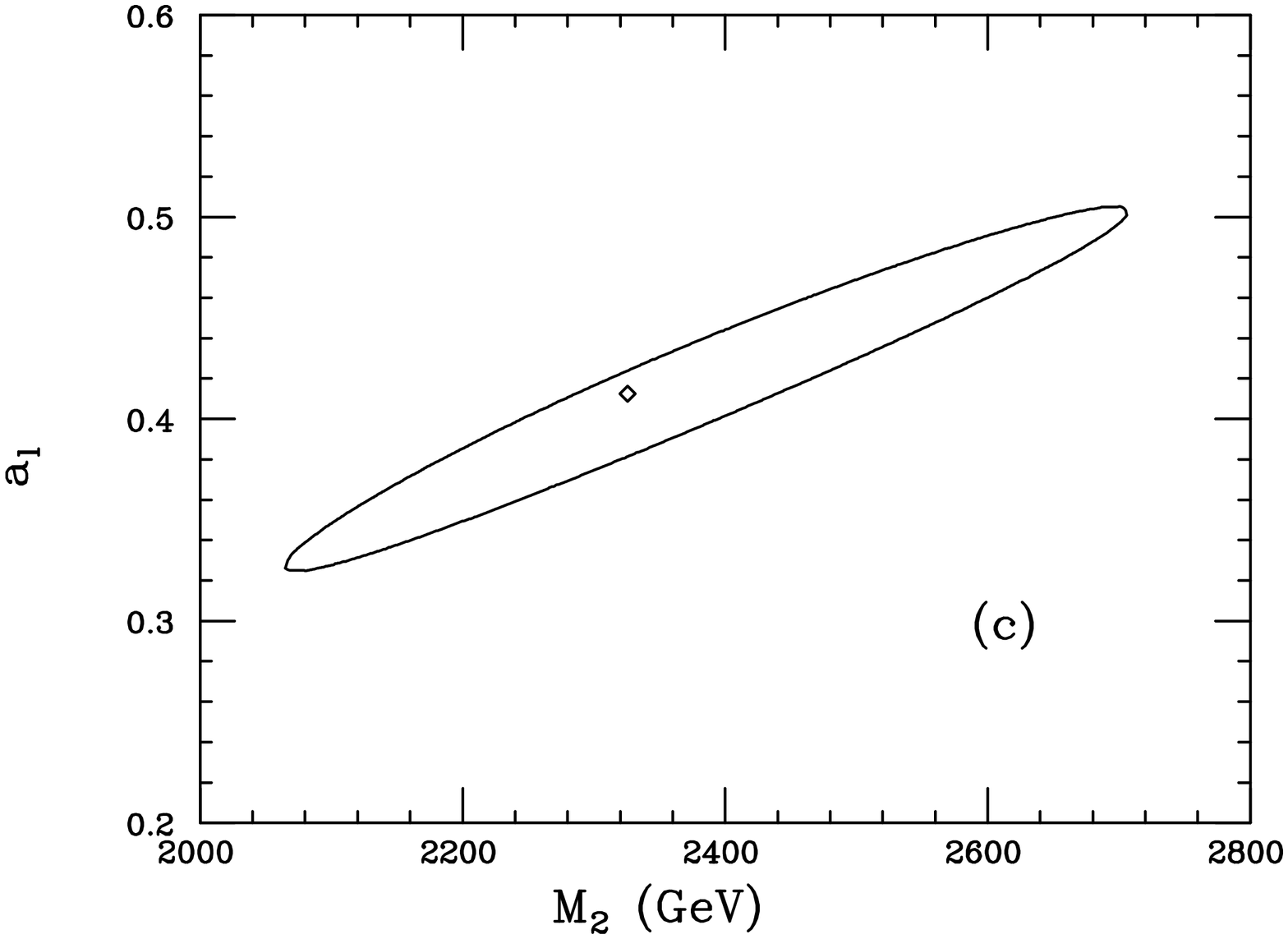,height=9.1cm,width=9.1cm,angle=-90}
\hspace*{-5mm}
\psfig{figure=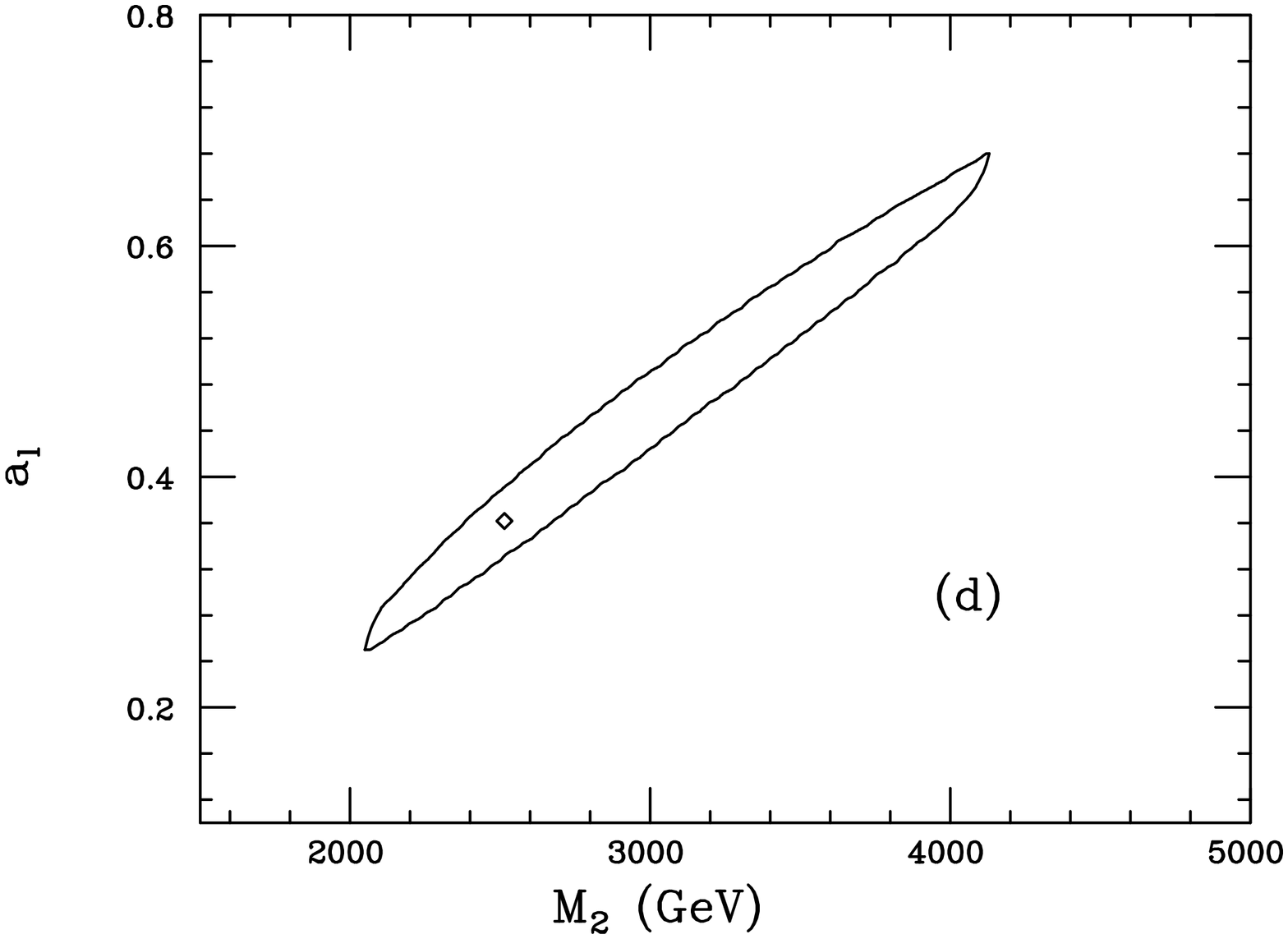,height=9.1cm,width=9.1cm,angle=-90}}
\vspace*{-0.3cm}
\caption{\small Lepton 
coupling determination for $Z'$'s with masses of (a) 2.33 TeV and (b) 2.51 TeV 
when the mass is unknown(solid) and known(dotted). (c) and (d) are the 
corresponding mass determinations which result from the five-dimensional 
fit. These results include an 
additional 200 $fb^{-1}$ of luminosity taken at a center of mass energy of 
1.5 TeV.} 
\end{figure}
\vspace*{0.4mm}

\Acknowledgements

The author would like to thank J.L. Hewett, S. Godfrey, K. Maeshima, and 
S. Riemann for discussions related to this work.

%
\def\MPL #1 #2 #3 {Mod.~Phys.~Lett.~{\bf#1},\ #2 (#3)}
\def\NPB #1 #2 #3 {Nucl.~Phys.~{\bf#1},\ #2 (#3)}
\def\PLB #1 #2 #3 {Phys.~Lett.~{\bf#1},\ #2 (#3)}
\def\PR #1 #2 #3 {Phys.~Rep.~{\bf#1},\ #2 (#3)}
\def\PRD #1 #2 #3 {Phys.~Rev.~{\bf#1},\ #2 (#3)}
\def\PRL #1 #2 #3 {Phys.~Rev.~Lett.~{\bf#1},\ #2 (#3)}
\def\RMP #1 #2 #3 {Rev.~Mod.~Phys.~{\bf#1},\ #2 (#3)}
\def\ZP #1 #2 #3 {Z.~Phys.~{\bf#1},\ #2 (#3)}
\def\IJMP #1 #2 #3 {Int.~J.~Mod.~Phys.~{\bf#1},\ #2 (#3)}

\end{document}